\documentclass[12pt]{article}

\begin{document}
\begin{flushright}
UPR-1165-T\\
SISSA-67/2006/EP\\
{\tt hep-th/0611128}
\end{flushright}
\vskip0.1truecm

\begin{center}
\vskip 2.5truecm {\Large \textbf{
(Non)-supersymmetric Marginal Deformations
from Twistor String Theory}}
\vskip 1truecm

{\large \textbf{Peng Gao${}^\star${}, Jun-Bao Wu${}^\S$}}\\

\vskip .7truecm
$^\star${\it Department of Physics and Astronomy\\
University of Pennsylvania\\
Philadelphia, PA 19104-6396, USA}\\
\tt{gaopeng@physics.upenn.edu}
\vskip 4truemm ${}^\S${\it International School for Advanced Studies (SISSA)\\
via Beirut 2-4, I-34014 Trieste, ITALY}\\
\tt{wujunbao@sissa.it}
\vskip 4truemm

\end{center}

\vskip 2truecm

\begin{center}

\textbf{\large Abstract}

\end{center}

The tree-level amplitudes in $\beta$-deformed theory are studied
from twistor string theory. We first show that a simple
generalization of the proposal in hep-th/0410122 gives the correct
results for all of the tree-level amplitudes to the first order of
the deformation parameter $\beta$. Then we give a proposal to all
orders of $\beta$ and show this matches the field theory. We also
show the prescription using connected instantons and the
prescription using disconnected instantons are equivalent in the
deformed twistor string theory. The tree-level amplitudes in
non-supersymmetric $\gamma$-deformed theory are also obtained in
this framework. The tree-level purely gluonic amplitudes in
theories with generic marginal deformations are also discussed in
the twistor string theory side.

\section{Introduction}

One of the important issues in AdS/CFT correspondence \cite{Maldacena:1997re,
Witten:1998qj, Gubser:1998bc} is to study the gravity dual of the gauge
theories with less supersymmtries. In \cite{Lunin:2005jy}, the gravity dual of
some
gauge theories with less supersymmtries was studied using a solution generating
transformation. One of the gauge theories studied in \cite{Lunin:2005jy} is the
$\beta$-deformed theory which has ${\cal N}=1$ supersymmetry. Later this
discussion was generalized to $\gamma$-deformed theory which has no
supersymmtries \cite{Frolov:2005dj, Frolov:2005ty,
Durnford:2006nb}.\footnote{The deformations in gauge theory were also studied
in \cite{Beisert:2005if}-\cite{Chu:2006ae} and some generalizations of the work in \cite{Frolov:2005dj}
including a generalization of
$\gamma$-deformation can be found in \cite{Alday:2005ww}.}
These deformed SYM can be obtained from a certain kind of star product among the superfields. Using this star product, it is easy to see that the planar amplitudes in such
theories are identical to those of the ${\cal N}=4$ theory up to an overall phase factor \cite{Lunin:2005jy, Khoze:2005nd, Durnford:2006nb}.

On the other hand, Witten found a new relation between
perturbative gauge theory and a topological string theory whose
target space is the supertwistor space \cite{Witten:2003nn}. While
the proposed AdS/CFT correspondence is a strong-weak duality, the
new correspondence found by Witten is a perturbative one. The
gauge theory amplitudes can then be computed using localization to either the connected
instantons in twistor space \cite{Witten:2003nn, Roiban:2004vt,
Roiban:2004ka, Roiban:2004yf}, or the completely disconnected instantons \cite{Cachazo:2004kj}.
 Later it was shown that in fact these two prescriptions are equivalent \cite{Gukov:2004ei}
and a family of intermediate prescriptions are also studied
\cite{Gukov:2004ei, Bena:2004ry}. Following Witten's original paper
\cite{Witten:2003nn}, there appeared a lot of works on computing tree-level
 and/or one-loop amplitudes in four dimensional gauge theories, some of which are \cite{Zhu:2004kr}-\cite{Xiao:2006vt}.

The twistor string theory corresponding to the gauge theories with
less supersymmetries obtained by the Leigh-Strassler deformation
\cite{Leigh:1995ep} was proposed in \cite{Kulaxizi:2004pa},
motivated by earlier work on deformations in the open-closed
topological string theories \cite{Hofman:2000ce, Hofman:2002cw}.
There it was suggested that the effect of the marginal
deformations in field theory can be captured in twistor string
theory by introducing a non-anticommutative star product among
the fermionic coordinates in the superstwistor space. This star product can be shown
heuristically to partially arise from a particular fermionic deformation in the closed string background.
A prescription for calculating the tree-level amplitudes in the
deformed theory using the connected instantons was also given in
that paper to the first order of the deformation parameters. In
this prescription, a non-anti-commutative star product among the
wavefunctions of the external particles plays an important role.
The authors of \cite{Kulaxizi:2004pa} computed examples of
amplitudes corresponding to degree one curves in twistor space
(these amplitudes are dubbed 'analytic amplitudes' in
\cite{Georgiou:2004wu, Georgiou:2004by}) and found that the
results coincide with field theory results. Later they claimed
one can generalize the prescription to amplitudes from higher
degree curves \cite{Kulaxizi:2005qt}. The precise origin of this
prescription is still not well understood. In particular, a
prescription valid to all orders of the deformation parameter is
absent until now. It is therefore interesting to make some
progress in this direction.

Since the planar amplitudes in $\beta$-deformed and
$\gamma$-deformed theories are rather simple, one may wonder
whether they can be reproduced using the twistor string theory. Up
to now it's not clear how to calculate the amplitudes beyond
tree-level from twistor string technique, we will from now on
focus on tree-level amplitudes. We first show that the tree-level
amplitudes in $\beta$-deformed theory  can be obtained from the
prescription in \cite{Kulaxizi:2004pa} to first order of the
deformation parameter. Notice our result is not restricted to the
degree one case. In fact, we study all of the tree-level
amplitudes exploiting a rather simple generalization of
\cite{Kulaxizi:2004pa} based on previous work for the ${\cal N}=4$
theory \cite{Roiban:2004vt, Roiban:2004ka, Roiban:2004yf}. We then
propose an all-order exact prescription and show that this new
prescription gives the right field theory results. The MHV
diagrams and the equivalence between the prescriptions using
connected instantons and completely disconnected instantons are
also discussed in the case of the $\beta$-deformed theory. The
$\gamma$-deformed theory is no longer supersymmetric, and doesn't
belong to the class of field theories studied in
\cite{Kulaxizi:2004pa}. In this case, we also give an prescription
to all orders of $\gamma$ to reproduce tree-level amplitudes from
the twistor string theory.

Encouraged by these results, we proceed to find some further
evidence for a possible all-degree generalization of
\cite{Kulaxizi:2004pa} for theories with generic
Leigh-Strassler deformations. We consider tree-level scattering
amplitudes involving only particles in the ${\cal N}=1$ vector
supermultiplet. One can easily see that the particles in the the
${\cal N}=1$ chiral supermultiplets cannot appear in the contributing
tree-level Feynman diagrams. So these amplitudes in the theory with
generic Leigh-Strassler deformations are the same as the ones in
the undeformed ${\cal N}=4$ theory. In \cite{Kulaxizi:2004pa}, this result was
shown to be valid for the analytic amplitudes  to the linear order of the
deformation paramters. Here we improve on this result by showing that it is in fact true for all
of the tree-level amplitudes involving only particles in ${\cal N}=1$
vector supermultiplet.

This paper is organized as follows. In the next section we briefly
review the method of computing field theory tree-level amplitudes from connected
instantons in twistor string theory. In section 3, we study the
amplitudes in $\beta$-deformed gauge theory from the twistor
string theory. An exact prescription for the amplitudes in
$\gamma$-deformed theory is given in section 4. In section 5, we
discuss the tree-level amplitudes with only particles in the
${\cal N}=1$ vector supermultiplet in the case of generic
Leigh-Strassler deformations. The final section devoted to
conclusions and discussions on further studies.

\section{Review of tree-level Yang-Mills amplitudes from connected instantons in
twistor string}
\subsection{Amplitudes of the ${\cal N}=4$ theory from twistor string theory}
We denote the tree-level partial amplitude of ${\cal N}=4$ super
Yang-Mills theory, which is the coefficient of
\begin{equation}
{\rm Tr}(T^{a_1}\cdots T^{a_n}),
\end{equation}
by \begin{equation} {\cal A}^{{\cal N}=4}_n(\pi_1, \tilde\pi_1,
h_1, t_1;\cdots; \pi_n, \tilde\pi_n, h_n, t_n),
\end{equation}
or simply by
 ${\cal A}^{{\cal N}=4}_n(\{\pi_i, \tilde\pi_i, h_i, t_i\})$, here
we decompose the momenta of the ${\rm i}$-th massless extrenal particles
into two Weyl spinors, ie., we define the 'bi-spinor'
$p_{ia\dot{a}}$ as $p_{ia\dot{a}}\equiv
p_{i\mu}\sigma^{\mu}_{a\dot{a}}$ and choose a decomposition of
this 'bi-spinor' as $p_{ia\dot{a}}=\pi_{ia}\tilde\pi_{i\dot{a}}$
(for more details, see, for example \cite{Witten:2003nn,
Cachazo:2005ga}), and we use $h_i$ to denote the helicity of this
external particle and $t_i$ its $SU(4)_R$ quantum numbers.

This amplitude can be computed in twistor string theory from localization on the
connected instantons \cite{Witten:2003nn, Roiban:2004vt,
Roiban:2004ka, Roiban:2004yf} (some aspects of this approach were
summarized in \cite{Cachazo:2005ga}). The main formula is:
\begin{equation}
{\cal A}^{{\cal N}=4}_n(\{\pi_i, \tilde\pi_i, h_i, t_i\})=\int
\mathrm{d}{\cal M}_d \langle\int_C J_1\Psi_1 \cdots \int_C
J_n\Psi_n\rangle, \label{maineq1}
\end{equation}
in the following we will explain the ingredients in this formula.
First we note that in the supertwistor space ${\bf CP}^{3|4}$ with
coordinates denoted by $Z^I=(\lambda^a, \mu^{\dot{a}}), \psi^A, (a,
\dot{a}=1,2, I, A=1,\cdots, 4)$, we have the following expansion
\cite{Witten:2003nn}:
\begin{eqnarray}
{\cal
A}&=&A+\psi^4\chi_4+\psi^I\chi_I+\psi^4\psi^I\phi_I+{1\over2}\psi^I\psi^J\epsilon_{IJK}\tilde\phi^K
+{1\over2}\psi^4\psi^I\psi^J\epsilon_{IJK}\tilde\chi^K\nonumber\\&+&{1\over
3!}\psi^I\psi^J\psi^K\epsilon_{IJK}\tilde\chi^4
+{1\over3!}\psi^4\psi^I\psi^J\psi^K\epsilon_{IJK}G.
\end{eqnarray}
As in \cite{Kulaxizi:2004pa}, we split the four fermionic
coordinates in ${\bf CP}^{3|4}$, $\psi^A (A=1, \cdots, 4)$, into
$\psi^I\, (I=1, 2, 3)$ and $\psi^4$. For latter convenience, we
write the above expansion as
\begin{equation}
 {\cal A}=\sum_{h=-1, -1/2, 0, 1/2, 1}g^a_h(\psi^4, \psi^I)\Phi_{ha}(\lambda, \mu),
\end{equation}
where $\Phi_{ha}=A, \chi_4, \chi_I, \phi_I, \tilde\phi^I,
\tilde\chi^K, \tilde\chi^4, G$ respectively. $\Psi_i(\pi_i,
\tilde\pi_i, h_i, t_i)$ in eq.~(\ref{maineq1}) is the wavefunction
of the ${\rm i}$-th particle in the supertwistor space, which is given by
\begin{equation}
 \Psi_i(\pi_i, \tilde\pi_i, h_i, t_i)=\bar{\delta}(\langle\lambda,
 \pi_i\rangle)\left({\lambda\over\pi_i}\right)^{2h-1}
\exp{\left(i[\tilde\pi_i, \mu](\pi_i/\lambda)\right)}g^a_h(\psi^4,
\psi^I),
\end{equation}
where the definition of the delta function $\bar\delta(f)$ is
\begin{equation}
 \bar\delta(f)=\bar \partial\bar f\delta^2(f)
\end{equation}
following \cite{Cachazo:2004kj}. In eq.~(\ref{maineq1}), $J_i$ is
a holomorphic current made of free fermions in the wouldvolume
theory of the $D5$-branes (the details can be found in
\cite{Witten:2003nn, Cachazo:2005ga}). Since what we actually
compute is the partial amplitude which is the coefficient of ${\rm
Tr}(T^{a_1}\cdots T^{a_n})$, in eq.~(\ref{maineq1}) we only pick
out the following coefficient of ${\rm Tr}(T^{a_1}\cdots T^{a_n})$
in the correlation function $\langle J_1(u_1)\cdots
J_n(u_n)\rangle$:
\begin{equation}
 {\prod_i\langle u_i, \mathrm{d}u_i \rangle\over\prod_{k}
\langle u_k, u_{k+1} \rangle}.
\end{equation}
In the prescription using the connected $D$-instantons which are
D-strings wrapped on algebraic curves, these amplitudes only
receive contributions from the curves with genus zero and
degree satisfying $d={1\over 2}\sum_{i=1}^n(1-h_i)-1$.

If we choose homogeneous coordinates $(u, v)$ on an abstract ${\bf
CP}^1$, then the genus 0, degree $d$ curve $C$, which is a map
from ${\bf CP}^1$ to ${\bf CP}^{3|4}$, can be parametrized as
\begin{eqnarray}
 Z^I&=&P^I(u, v)=\sum_{\alpha=0}^d
P^I_{\alpha}u^{\alpha}v^{d-\alpha},\nonumber\\
 \psi^A&=&Q^A(u, v)=\sum_{\alpha=0}^d Q^A_{\alpha}u^{\alpha}v^{d-\alpha}.
\end{eqnarray}
Then the measure on the moduli space of the genus 0, degree $d$
curves can be written as
\begin{equation}
 \mathrm{d}{\cal M}_d={\prod_{\alpha=1}^d\prod_{A=1}^4\prod_{I=1}^4
\mathrm{d}P^I_\alpha \mathrm{d}Q^A_\alpha\over GL(2, {\bf C})}.
\end{equation}

Now we may write eq.~(\ref{maineq1}) as
\begin{eqnarray}
{\cal A}^{{\cal N}=4}_n(\{\pi_i, \tilde\pi_i, h_i, t_i\})&=&\int
\mathrm{d}{\cal M}_d \prod_i \int_C \langle u_i, \mathrm{d}u_i
\rangle\bar\delta(\langle\lambda(u_i),
\pi_i\rangle)\left({\lambda(u_i)\over\pi_i}\right)^{2h_i-1}\nonumber\\
&\times&\exp{(i[\mu(u_i),
\tilde\pi_i](\pi_i/\lambda(u_i)))}g^{a_i}_{h_i}(\psi_i){1\over
\prod_k \langle
u_k, u_{k+1}\rangle}.\nonumber\\
\end{eqnarray}

\subsection{Amplitudes in deformed SYM from twistor string theory}
In \cite{Kulaxizi:2004pa}, a proposition was proposed to calculate
the amplitudes in deformed SYM from twistor string theory to the
linear order of the deformation parameters.

The superpotential in the deformed theory considered in
\cite{Kulaxizi:2004pa} can be written as\footnote{As in
\cite{Kulaxizi:2004pa}, we pull the gauge coupling constant out of
the superpotential. For ease of comparison, we also use the same normalization as
\cite{Kulaxizi:2004pa}.}
\begin{equation}
{\cal W}={\cal W}_{{\cal N}=4}+{1\over 3!}h^{IJK}{\rm
Tr}(\Phi_I\Phi_J\Phi_K),\label{deformation}
\end{equation}
to linear order of the deformation parameters $h^{IJK}$, where
$h^{IJK}$ is totally symmetric. Here
\begin{equation}
{\cal W}_{{\cal N}=4}=i{\rm Tr}(\Phi_1[\Phi_2,\Phi_3]),
\end{equation}
is the superpotential of the ${\cal N}=4$ SYM and $\Phi_i, i=1, 2,
3$ are three chiral superfields in the adjoint representation of
the gauge group. We note that the deformed theory only has ${\cal
N}=1$ supersymmetry.

The authors of \cite{Kulaxizi:2004pa} used a star product among
the $\psi$-dependent part of the wavefunctions (to the first order
of the deformation parameters) as follows \footnote{We note that
the superscript of $\psi$ denotes its $SU(4)$ R-symmetry quantum
number (the $SU(4)_R$ symmetry of the ${\cal N}=4$ SYM has been
broken by the deformation) and the subscript of $\psi$ numbers the
corresponding external particle.}
\begin{equation}
f(\psi_1)\ast g(\psi_2)=f(\psi_1)g(\psi_2)-{i \over 4} V^{IJ}_{KL}
\left(f(\psi_1)\overleftarrow{\frac{\partial}{\partial\psi^I_1}}\right)\psi^K_1
\psi^L_2\left(\overrightarrow{\frac{\partial}{\partial\psi^J_2}}g(\psi_2)\right),
\end{equation}
where the definition of $V^{IJ}_{KL}$ is
\begin{equation}
V^{IJ}_{KL}=h^{IJQ}\epsilon_{QKL}+\epsilon^{IJQ}{\bar h}_{QKL},
\end{equation}
then the amplitudes is given by
\begin{equation}
{\cal A}^{deformed}_n(\{\pi_i, \tilde\pi_i, h_i, t_i\})=\int
\mathrm{d}{\cal M}_d \langle\int_C J_1\Psi_1 \ast\int_C J_2\Psi_2
\ast \cdots \ast \int_C J_n\Psi_n\rangle,\label{eqdef}
\end{equation} to first order of the deformation parameters\footnote{As we have
mentioned in the introduction, in \cite{Kulaxizi:2004pa} only the
analytic amplitudes were discussed, the formula here is valid for
all tree-level amplitudes.}.

\section{Tree-level amplitudes in $\beta$-deformed theory from twistor string
theory}
\subsection{All tree-level amplitudes to linear order in $\beta$}
The superpotential in the $\beta$-deformed SYM theory
\cite{Lunin:2005jy, Khoze:2005nd} is
\begin{equation}
{\cal W}_{\beta-deformed}=i{\rm
Tr}(e^{i\pi\beta}\Phi_1\Phi_2\Phi_3-e^{-i\pi\beta}\Phi_1\Phi_3\Phi_2).
\end{equation}
Here $\beta$ is a real deformation parameter.

Now we compute the tree-level amplitudes in this theory using the
all-degree generalized prescription we give above. First, we
expand ${\cal W}_{\beta-deformed}$ to first order of $\beta$:
\begin{equation}
{\cal W}_{\beta-deformed}={\cal W}_{{\cal N}=4}-\pi\beta{\rm
Tr}(\Phi_1\{\Phi_2,\Phi_3\})+O(\beta^2).
\end{equation}
Then one can easily read off in this case the  $h^{IJK}$'s in
eq.~(\ref{deformation}) are simply $-2\pi\beta|\epsilon_{IJK}|$,
so we find
\begin{equation}
V^{IJ}_{KL}=h^{IJQ}\epsilon_{QKL}+\epsilon^{IJQ}{\bar h}_{QKL}
=-4\pi\beta\delta^I_K\delta^J_L\epsilon_{IJQ}\alpha_Q
\end{equation}
where the definition of $\alpha_I$'s is
$\alpha_1=\alpha_2=\alpha_3=1$.

Using this, it can be easily found that for arbitary wavefunction $f$
and $g$ we have
\begin{eqnarray}
f(\psi_1)\ast g(\psi_2)&=&f(\psi_1)g(\psi_2)+i\pi\beta
f(\psi_1)\left(\sum_{I, J,
K}\epsilon_{IJK}\overleftarrow{\frac{\partial}{\partial
\psi^I_1}}\psi^I_1\alpha_K\psi^J_2\overrightarrow{\frac{\partial}{\partial
\psi^J_2}}\right)g(\psi_2)\nonumber\\&+&O(\beta^2).\label{eqbeta1}
\end{eqnarray}
Because the wavefunctions are all monomials in fermionic directions, we further have
\begin{equation}
f(\psi_1) \overleftarrow{\frac{\partial}{\partial \psi^I_1}}
\psi^I_1 =\left\{
\begin{array}{cc}
f(\psi_1) &\mbox{if $I \in f$,} \\
0         &\mbox{otherwise.}
\end{array}
\right.
\end{equation}
were by $I\in f$, we simply meant $\psi^I_1$ is a factor of $f(\psi_1)$.
In other words,
\begin{equation}
f(\psi_1) \overleftarrow{\frac{\partial}{\partial \psi^I_1}}
\psi^I_1 =N^I_f f(\psi_1),\end{equation} where $N^I_f$ is the
number of $\psi^I_1$ in $f(\psi_1)$ ($N^I_f \leq 1$ since
$\psi^I_1$ is fermionic). Similarly,
\begin{equation}
\psi^J_2 \overrightarrow{\frac{\partial}{\partial
\psi^J_2}}g(\psi_2)=N^J_g g(\psi_2).
\end{equation}
So eq.~(\ref{eqbeta1}) can also be re-written as
\begin{eqnarray}
f(\psi_1)\ast
g(\psi_2)&=&(1+i\pi\beta\sum_{I, J, K}\epsilon_{IJK}N^I_fN^J_g\alpha_K)f(\psi_1)g(\psi_2)+O(\beta^2)\nonumber\\
&=&(1+i\pi\beta\sum_{I\in f,J\in
g}\epsilon_{IJK}\alpha_K)f(\psi_1)g(\psi_2)+O(\beta^2),
\end{eqnarray}

We can slightly rephrase the above results by considering the following
$U(1)\times U(1)$ symmetry acting on the fermionic part of the
supertwistor space ${\bf CP}^{3|4}$, under which $(\psi^1, \psi^2,
\psi^3, \psi^4)$ are assigned the following charges \footnote{In
\cite{Ananth:2006ac}, similar star product was introduced in
${\cal N}=4$ light-cone superspace in stead of supertwistor space.
Recently, a possible new relation between $\beta$-deformation and
noncommutative field theories was discussed in
\cite{Kulaxizi:2006pp}.}:
\begin{equation}
\begin{array}{ccccc}
  & \psi^1 &\psi^2 &\psi^3 &\psi^4 \\
\tilde Q_1&0&-1&1&0\\
\tilde Q_2&1&-1&0&0
\end{array}
\end{equation}
from which we have $\epsilon_{IJK}\alpha_K=\tilde Q^{\psi^I}_1\tilde
Q^{\psi^J}_2-\tilde Q^{\psi^I}_2\tilde Q^{\psi^J}_1$. Using the identity
\begin{equation}
\tilde Q(\psi^{I_1}\cdots\psi^{I_n})=\tilde Q^{\psi^{I_1}}+\cdots
\tilde Q^{\psi^{I_n}},
\end{equation}
we get
\begin{equation}
\Psi_1\ast\Psi_2\ast\cdots\ast\Psi_n=\left(1+i\pi\beta\sum_{i<j}(\tilde
Q^i_1\tilde Q^j_2-\tilde Q^i_2\tilde
Q^j_1)\right)\Psi_1\Psi_2\cdots\Psi_n+O(\beta^2),
\end{equation}
Here $\tilde Q^i_1$ and $\tilde Q^i_2$ are the charges of the
${\rm i}$-th wavefunction. We notice that there is the following relation
between the $\tilde Q_i$ defined here and the $Q_i$ defined in
\cite{Lunin:2005jy, Khoze:2005nd},
\begin{equation}
Q_i(\Phi_{ha})+\tilde Q_i(g^a_h(\psi^4, \psi^I))=0.\label{charges}
\end{equation}
Replacing the charges using this identification, we find
\begin{equation}
{\cal A}^{\beta-deformed}_n=\left(1+i\pi\beta\sum_{i<j}( Q^i_1
Q^j_2-Q^i_2 Q^j_1)\right){\cal A}^{{\cal N}=4}_n+O(\beta^2).
\end{equation}
Here $Q^i_1$ and $Q^i_2$ are the charges of the wavefunction of
$i$-th external particle. This result coincides with the results
in \cite{Khoze:2005nd} to first order of $\beta$.

\subsection{The star product to all orders of $\beta$}
Based on the calculation in the previous subsection, we propose the following star product among the wave functions
\begin{eqnarray}
f(\psi_1)\ast g(\psi_2)&=& f(\psi_1)\exp\left(i\pi\beta\sum_{I, J,
K}\epsilon_{IJK}\overleftarrow{\frac{\partial}{\partial
\psi^I_1}}\psi^I_1\alpha_K\psi^J_2\overrightarrow{\frac{\partial}{\partial
\psi^J_2}}\right)g(\psi_2)\nonumber\\
&=&\sum_{n=0}^{\infty}{(i\pi\beta)^n\over
n!}\sum_{I_1J_1K_1}\cdots\sum_{I_nJ_nK_n}f(\psi_1)(\epsilon_{I_1J_1K_1}\overleftarrow{\frac{\partial}{\partial
\psi^{I_1}_1}}\psi^{I_1}_1\alpha_{K_1}\psi^{J_1}_2\overrightarrow{\frac{\partial}{\partial
\psi^{J_1}_2}})\nonumber\\
&&\cdots(\epsilon_{I_nJ_nK_n}\overleftarrow{\frac{\partial}{\partial
\psi^{I_n}_1}}\psi^{I_n}_1\alpha_{K_n}\psi^{J_n}_2\overrightarrow{\frac{\partial}{\partial
\psi^{J_n}_2}})g(\psi_2)
\end{eqnarray}
 will reproduce all tree-level field theory amplitudes to all orders of $\beta$.

Since $\overleftarrow{\frac{\partial}{\partial
\psi^{I_l}_1}}\psi^{I_l}_1$ commutes with
$\psi^{J_m}_2\overrightarrow{\frac{\partial}{\partial
\psi^{J_m}_2}}$, we have
\begin{eqnarray}
f(\psi_1)\ast g(\psi_2) &=&\sum_{n=0}^{\infty}{(i\pi\beta)^n\over
n!}\sum_{I_1J_1K_1}\cdots\sum_{I_nJ_nK_n}\epsilon_{I_1J_1K_1}\cdots\epsilon_{I_nJ_nK_n}\alpha_{K_1}\cdots\alpha_{K_n}
\nonumber\\
&\times&f(\psi_1)\overleftarrow{\frac{\partial}{\partial
\psi^{I_1}_1}}\psi^{I_1}_1\cdots\overleftarrow{\frac{\partial}{\partial
\psi^{I_n}_1}}\psi^{I_n}_1
\psi^{J_1}_2\overrightarrow{\frac{\partial}{\partial
\psi^{J_1}_2}}\cdots\psi^{J_n}_2\overrightarrow{\frac{\partial}{\partial
\psi^{J_n}_2}}g(\psi_2)\nonumber\\
&=&\sum_{n=0}^{\infty}{(i\pi\beta)^n\over
n!}\sum_{I_1J_1K_1}\cdots\sum_{I_nJ_nK_n}\epsilon_{I_1J_1K_1}\cdots\epsilon_{I_nJ_nK_n}\alpha_{K_1}\cdots\alpha_{K_n}
\nonumber\\
&\times& N^{I_1}_f\cdots N^{I_n}_f N^{J_1}_g\cdots N^{J_n}_g
f(\psi_1)g(\psi_2) \nonumber\\
&=&\sum_{n=0}^{\infty}{(i\pi\beta)^n\over
n!}\left(\sum_{IJK}\epsilon_{IJK}\alpha_{K}N^{I}_f
N^{J}_g\right)^n f(\psi_1)g(\psi_2)\nonumber\\
&=&\sum_{n=0}^{\infty}{(i\pi\beta)^n\over n!}\left(\sum_{I\in
f,J\in g}\epsilon_{IJK}\alpha_K\right)^n f(\psi_1)g(\psi_2)
\nonumber\\
&=&\exp\left(i\pi\beta\sum_{I\in f,J\in
g}\epsilon_{IJK}\alpha_K\right)f(\psi_1)g(\psi_2)~.
\end{eqnarray}
As in the previous subsection, we can now re-write
\begin{equation}
f(\psi_1)\ast g(\psi_2)=\exp\left(i\pi\beta(\tilde Q_1^f \tilde
Q_2^g-\tilde Q_2^f \tilde Q_1^g)\right)f(\psi_1)g(\psi_2).
\end{equation}
From eq.~(\ref{charges}), we know that the star product among the
wavefunctions defined here is isomorphic to the one among the
fields defined in \cite{Lunin:2005jy, Khoze:2005nd}, and hence the former
star product is also associative due to the associativity of the latter one. This can be verified explicitly as follows:
\begin{eqnarray}
(f\ast g)\ast h&=&\exp\left(i\pi\beta(\tilde Q_1^f \tilde
Q_2^g-\tilde Q_2^f \tilde Q_1^g)\right) (fg)\ast h\nonumber\\
&=&\exp\left(i\pi\beta(\tilde Q_1^f \tilde Q_2^g-\tilde Q_2^f
\tilde Q_1^g)+(i\pi\beta(\tilde Q_1^{fg} \tilde Q_2^h-\tilde
Q_2^{fg} \tilde Q_1^h)\right)fgh\nonumber\\
&=& \exp\left(i\pi\beta(\tilde Q_1^f \tilde Q_2^{gh}-\tilde Q_2^f
\tilde Q_1^{gh})+(i\pi\beta(\tilde Q_1^g \tilde Q_2^h-\tilde Q_2^g
\tilde Q_1^h)\right)fgh\nonumber\\
&=& \exp\left(i\pi\beta(\tilde Q_1^g \tilde Q_2^h-\tilde Q_2^g
\tilde Q_1^h\right)f\ast(gh)\nonumber\\
&=& f\ast(g\ast h),\end{eqnarray} where $\tilde Q^{fg}_i=\tilde
Q^f_i+\tilde Q^g_i (i=1, 2)$ is used.

Similar to what we have done in the previous subsection, one can
easily verify that
\begin{equation}
{\cal A}^{\beta-deformed}_n=\exp\left(i\pi\beta\sum_{i<j}( Q^i_1
Q^j_2-Q^i_2 Q^j_1)\right){\cal A}^{{\cal N}=4}_n.
\end{equation}
This result is indeed the same as the one obtained in field theory computations \cite{Lunin:2005jy, Khoze:2005nd}.

\subsection{The MHV diagrams and the equivalence between different twistor space
 prescriptions}
In this subsection we will discuss the MHV diagrams and the equivalence between
the
prescription using connected instantons and the prescription using completely
disconnected instantons for the $\beta$-deformed theory\footnote{In this
subsection, we will only give some brief discussions, the
details are omitted since they are similar to the discussions in the ${\cal N}=4$
case.}.

The analytic amplitude in $\beta$-deformed theory is independent of
$\tilde\lambda$, since this amplitude equals the product of the corresponding
analytic amplitude in ${\cal N}=4$ theory and a phase factor\footnote{Notice
that the phase
factor also depends on the internal particles.}, both of which are
independent of $\tilde\lambda$. Then in the $\beta$-deformed theory, we can use
the same off-shell continuation as in \cite{Cachazo:2004kj, Georgiou:2004by} to
define the analytic vertex. We can compute all of the tree-level amplitudes in
$\beta$-deformed theory from the MHV diagrams obtained by connecting the
analytic vertices with
propagators as in \cite{Georgiou:2004by}. Similar to the proof using Feynman
rules in
\cite{Khoze:2005nd}, we can prove that the amplitudes from the MHV diagrams are
exactly the same as expected.

As in the ${\cal N}=4$ theory  \cite{Cachazo:2004kj}, the
extended-CSW rules in the $\beta$-deformed theory mentioned above
can be obtained from the prescription for calculating the
tree-level amplitudes using completely disconnected instantons in
the twistor string theory. Consider an expression in the
completely disconnected prescription corresponding to a given MHV
diagram. Notice that the twistor space propagator $D$ propotional
to $\delta^4(\psi-\psi^\prime)$ and the latter is just
$(\psi-\psi^\prime)^4$ \cite{Gukov:2004ei, Cachazo:2005ga} ($\psi$
and $\psi^\prime$ are fermionic coordinates of the two ends of the
twistor space propagator). In a given MHV diagram, each end of a
propagator has a fixed helicity and a fixed $SU(4)_R$ quantum
number. According to this helicity and quantum number, only one
term in the expansion of $(\psi-\psi^\prime)^4$ will be picked
out. This term is a product of two factors, one factor is a
product of $\psi$'s the other of $\psi^\prime$'s. So for every
analytic vertex and a propagator connected to this vertex, there
is an end of the propagator corresponding to the vertex. Then
there is a corresponding product of the fermionic coordinates. In
the expression corresponding to the MHV diagram, for every vertex,
these factors from the propagators connected to it and the
wavefunctions of the external particles connected to it should be
multiplied using the star product in supertwistor space to give
the amplitudes in $\beta$-deformed theory. Then, similar to the
proof using Feynman rules in \cite{Khoze:2005nd}, we will finally
get a star product among all of the external wavefunctions
multiplied by ordinary multiplication with the ordinary product of
the twistor space propagators. The associativity of the all-order
star product defined in the last subsection is essential for the
these arguments.

Following the proof in \cite{Gukov:2004ei} for the ${\cal N}=4$ case, one can
prove that the prescription using the connected instantons and the prescription using the
commpletely disconnected instantons will give the same result. One need only
notice that in the proof in \cite{Gukov:2004ei}, the poles of the integrands
which contribute to the residues never came from the wave-functions themselves, nor from the case when two points where the wavefunctions are inserted come close to each other.

\section{Tree-level amplitudes in $\gamma$-deformed theory
from twistor string theory}
A generalization of the $\beta$-deformation is the $\gamma$-deformation
\cite{Frolov:2005dj, Frolov:2005ty, Durnford:2006nb}. It is obtained by
introducing the following star product
\begin{equation}
f\star g=\exp\left(-i\pi Q_i^f Q_j^g
\epsilon_{ijk}\gamma_k\right)fg\end{equation}
among the component fields in the Lagrangian of ${\cal N}=4$ theory.
Here $\gamma_i, i=1, 2, 3$ are three real deformation parameters and $Q_i, i=1,
2, 3$ are the charge of three
$U(1)$ symmetries of the theory. The charges of the component fields are the
following \cite{Durnford:2006nb}:
\begin{equation}
\begin{array}{ccccccccc}
 &A_\mu&\phi_1&\phi_2&\phi_3&\chi_1 &\chi_2 &\chi_3 &\chi_4 \\
 Q_1&0&1&0&0&1/2&-1/2&-1/2&1/2\\
Q_2&0&0&1&0&-1/2&1/2&-1/2&1/2\\
Q_3&0&0&0&1&-1/2&-1/2&1/2&1/2
\end{array}
\end{equation}
and for every component field $\phi$, we have $Q_i(\phi^\dagger)=-Q_i(\phi)$.

The $\gamma$-deformed theory is non-supersymmetric and the component Lagrangian
of this theory can be found for example in \cite{Durnford:2006nb}.
To reproduce the tree-level amplitudes in the $\gamma$-deformed theory
\cite{Frolov:2005dj, Frolov:2005ty, Durnford:2006nb} from connected instantons
in twistor string theory, we consider the following global $U(1)^3$ symmetry
acting on the fermionic coordinates of the supertwistor space,
under which $(\psi^1, \psi^2, \psi^3, \psi^4)$ have the following charge assignment
\begin{equation}
\begin{array}{ccccc}
  & \psi^1 &\psi^2 &\psi^3 &\psi^4 \\
\tilde Q_1&-1/2&1/2&1/2&-1/2\\
\tilde Q_2&1/2&-1/2&1/2&-1/2\\
\tilde Q_3&1/2&1/2&-1/2&-1/2
\end{array}
\end{equation}
and we define the star product among the
wave functions as

\begin{equation}
f(\psi_1)\ast g(\psi_2)=f(\psi_1)\exp\left(-i\pi\sum_{A,
B}\overleftarrow{\frac{\partial}{\partial \psi^A_1}}\psi^A_1\tilde Q_i^{\psi^A}
 \epsilon_{ijk}\gamma_k\tilde Q_j^{\psi^B}\psi^B_2
\overrightarrow{\frac{\partial}{\partial \psi^B_2}}\right)g(\psi_2),
\end{equation}
which is equivalent to
\begin{equation}
f(\psi_1)\ast g(\psi_2)=\exp\left(-i\pi\sum_{A\in f,B\in
g}\tilde Q_i^{\psi^A} \tilde Q_j^{\psi^B}
\epsilon_{ijk}\gamma_k\right)f(\psi_1)g(\psi_2).
\end{equation}
This star product is also associative and it gives rise to the following
tree-level amplitudes:
\begin{eqnarray}
{\cal A}^{\gamma-deformed}_n&=&\exp\left(-i\pi\sum_{a<b}\tilde Q_i^a \tilde
Q_j^b \epsilon_{ijk}\gamma_k\right){\cal A}^{{\cal N}=4}_n\nonumber\\
&=&\exp\left(-i\pi\sum_{a<b}Q_i^a Q_j^b \epsilon_{ijk}\gamma_k\right){\cal
A}^{{\cal N}=4}_n,
\label{last}\end{eqnarray}
again coinciding with field theory results \cite{Durnford:2006nb} to all orders of
$\gamma$. Similarly to what we have done in the previous section, we can show
that the prescription using the connected instantons and the one using the
completely disconnected instantons are equivalent which relies crucially on the associativity of the star product just defined.

\section{Tree-level purely gluonic amplitudes in the general
deformed theory}

The tree-level purely gluonic amplitudes in theories with generic
Leigh-Strassler deformations are the same as in the ${\cal N}=4$
theory \cite{Kulaxizi:2004pa}. This is a trivial result in the
field theory side since if the external particles are all gluons,
the internal particles in the tree-level Feynman diagrams can only
be gluons too. In the twistor string theory side, this was proved
for the analytic amplitudes to first order of the deformation
parameters in \cite{Kulaxizi:2004pa}. Here we will show that it is
also true for all of the tree-lvel amplitudes obtained from
twistor string theory. We take this as another consistent check of
the all-degree generalization of the their prescription, ie.,
eq.~(\ref{eqdef}). We only consider the amplitudes $A_n(1^-, 2^-,
3^-, 4^+, \cdots, n^+)$ as an example. The demonstration of this
result for other purely gluonic amplitudes is identical in spirit.

The star product one needs to compute is
\begin{eqnarray}
&&(\epsilon_{I_1I_2I_3}\psi_1^{I_1}\psi_1^{I_2}\psi_1^{I_3})\ast
(\epsilon_{J_1J_2J_3}\psi_2^{J_1}\psi_2^{J_2}\psi_2^{J_3})\ast
(\epsilon_{K_1K_2K_3}\psi_3^{K_1}\psi_3^{K_2}\psi_3^{K_3})\nonumber\\
&=& \epsilon_{I_1I_2I_3}\epsilon_{J_1J_2J_3}\epsilon_{K_1K_2K_3}
\left(\psi_1^{I_1}\psi_1^{I_2}\psi_1^{I_3}\psi_2^{J_1}\psi_2^{J_2}\psi_2^{J_3}\psi_3^{K_1}\psi_3^{K_2}\psi_3^{K_3}\right.\nonumber \\
&-&{i\over 4} (3^2\,\psi_1^{I_1}\psi_1^{I_2}\psi_1^{\tilde I_3}
\psi_2^{\tilde
J_1}\psi_2^{J_2}\psi_2^{J_3}\psi_3^{K_1}\psi_3^{K_2}\psi_3^{K_3}
V^{I_3J_1}_{\tilde I_3 \tilde J_1}\nonumber\\&+&
3^2\,\psi_1^{I_1}\psi_1^{I_2}\psi_1^{I_3}
\psi_2^{J_1}\psi_2^{J_2}\psi_2^{\tilde J_3}\psi_3^{\tilde
K_1}\psi_3^{K_2}\psi_3^{K_3} V^{J_3K_1}_{\tilde J_3 \tilde
K_1}\nonumber\\&+&\left.
3^2\,\psi_1^{I_1}\psi_1^{I_2}\psi_1^{\tilde I_3}
\psi_2^{J_1}\psi_2^{J_2}\psi_2^{J_3}\psi_3^{\tilde
K_1}\psi_3^{K_2}\psi_3^{K_3} V^{I_3K_1}_{\tilde I_3 \tilde
K_1})\right),\label{pure}
\end{eqnarray}
 to the linear order of deformation.

The first term in the right hand side of the above equation
gives the amplitudes in ${\cal N}=4$ theory  \cite{Witten:2003nn,
Roiban:2004vt, Roiban:2004ka, Roiban:2004yf, Gukov:2004ei}. Now we
will show that the second term gives no contributions.
Since there are $3$ external gluons with negative helicity, the
contributing algebraic curves in supertwistor space should be the
ones with genus zero and degree $2$. Such curves can be parametrized as
\begin{eqnarray}
Z^I&=&\sum_{\alpha=0}^2 P^I_\alpha u^\alpha v^{2-\alpha},\\
\psi^A&=&\sum_{\alpha=0}^2 Q^A_\alpha u^\alpha v^{2-\alpha},
\end{eqnarray}

Now we define
\begin{equation}
F^{I_1I_2I_3J_1J_2J_3K_1K_2K_3}\equiv \int \prod_{I=1}^3
\left(\prod_{\alpha=0}^2d\,Q_\alpha^I\right)
\psi_1^{I_1}\psi_1^{I_2}\psi_1^{I_3}\psi_2^{J_1}\psi_2^{J_2}\psi_2^{J_3}\psi_3^{K_1}\psi_3^{K_2}\psi_3^{K_3}.
\end{equation}
It is not hard to see that \cite{Roiban:2004vt}
\begin{equation}
F^{I_1I_2I_3J_1J_2J_3K_1K_2K_3} \propto
\epsilon^{I_1I_2I_3}\epsilon^{J_1J_2J_3}\epsilon^{K_1K_2K_3},\label{eqf}
\end{equation}
here the proportional coefficient is independent of the indices
$I_i, J_i, K_i$ (we will not need the concrete value of this
coefficient in the following).

From this result, we can prove that
\begin{equation}
\epsilon_{I_1I_2I_3}\epsilon_{J_1J_2J_3}\epsilon_{K_1K_2K_3}
F^{I_1I_2\tilde I_3\tilde J_1J_2J_3K_1K_2K_3}V^{I_3J_1}_{\tilde
I_3 \tilde J_1 }=0.\end{equation}

Using eq.~(\ref{eqf}), we know that the left hand side of the
above equation is proportional to
\begin{eqnarray}
& &\epsilon_{I_1I_2I_3}\epsilon_{J_1J_2J_3}\epsilon_{K_1K_2K_3}
\epsilon^{I_1I_2\tilde I_3}\epsilon^{\tilde
J_1J_2J_3}\epsilon^{K_1K_2K_3}(\epsilon^{I_3J_1Q}{\bar h}_{Q\tilde
I_3 \tilde J_1}+h^{I_3J_1Q}\epsilon_{Q\tilde I_3 \tilde
J_1})\nonumber\\&\propto&\delta^{\tilde I_3}_{I_3}\delta^{\tilde
J_1}_{J_1}(\epsilon^{I_3J_1Q}{\bar h}_{Q\tilde I_3 \tilde
J_1}+h^{I_3J_1Q}\epsilon_{Q\tilde I_3 \tilde
J_1})=\epsilon^{I_3J_1Q}{\bar
h}_{QI_3J_1}+h^{I_3J_1Q}\epsilon_{QI_3J_1}\nonumber\\&=&0.\end{eqnarray}

So the second term in eq.~(\ref{pure}) vanishes after integral
over the Grassmann odd coordinates of the moduli space of these
curves. By the same calculation we find that neither the third
term nor the forth term gives contributions. This completes our
proof.

Using the same method, we can show that all amplitudes with only gluons and/or
gluinos in the ${\cal N}=1$ vector supermultiplet ( ie.,
$\lambda_4$) in the deformed theory are the same as in the ${\cal
N}=4$ theory to first order of deformations considered here.

\section{Conclusion and discussions}
In this paper, we studied the particle scattering amplitudes in four dimensional gauge theories with marginal deformations using the recently proposed twistor string theory, especially in the supersymmetric $\beta$-deformed theory and the non-supersymmetric $\gamma$-deformed theory.

In the $\beta$-deformed theory, we first re-investigate the amplitudes
to linear order in $\beta$ using a generalization of the prescription in \cite{Kulaxizi:2004pa}.
We find that the corresponding star product among the wavefuctions in this special case is drastically simplified compared to theories with general deformations. The special combinations $\overleftarrow{\frac{\partial}{\partial \psi^{I}_1}}\psi^{I}_1$ and $\psi^{J}_2\overrightarrow{\frac{\partial}{\partial\psi^{J}_2}}$ that appear in our expressions make it easy to
convert our star-product into the form found in \cite{Lunin:2005jy, Khoze:2005nd}. These special operators merely count the number of $\psi^I_1$ and $\psi^J_2$ and this fact lead us to make a conjecture for the all-order star product and show that this conjecture is indeed correct. In other words,
this identifies the exact star product we need to use to multiply two general functions $f(\psi_1)$ and $g(\psi_2)$ in the corresponding twistor string theory.
In sharp contrast to our success in the real $\beta$ case, it turned out much harder to find the exact star product in other more general cases at this moment. We hope to return to this problem in the near future.

In the case of real $\gamma$ deformed theory, we were also able to find an
all-order description which leads correctly to the field theory amplitudes.
This is done essentially by by guesswork and we know even less about aspects of the B-model description in this case. It is an interesting open problem to find the correct closed string background
which would give the self-dual part of the Lagrangian of this theory.

Finally we showed that the amplitudes involving only the fields in the
${\cal N}=1$ vector supermultiplet obtained from the prescription
is the same as the one expected from the field theory. Due to the difficulty alluded to above in generalizing our results in the special cases, we do this to linear order in the deformation parameters. This general result may be taken as yet another evidence of the possibility of completely understanding the closed B-model string background for the general marginal deformation of four dimensional gauge theory.

Despite the relative ease with which we were able to reproduce correct field theory amplitudes, we still lack a proper understanding of the precise origin of the prescription proposed in
\cite{Kulaxizi:2004pa}.  Notice the difficulty here is closely related to the fact that the scattering amplitudes involve products among different wavefunctions while the holomorphic Chern-Simons theory concerns only the wavefunction of the five-brane. This means necessarily that modifying the product in the holomorphic Chern-Simons theory won't give rise to the full star product required to reproduce four dimensional field theory results. We hope that our exact results in the above special cases may shed some more light on this issue.

Another interesting problem is to prove that the tree-level
amplitudes obtained from twistor string theory satisfy the
constraints from the parity invariance of the gauge theory. Since
in twistor theory, the gluons with opposite helicities are treated
in different manners, the parity invariance is not manifest at
all. It has been proven that the amplitudes in ${\cal N}=4$ super
Yang-Mills theory obtained from twistor string theory satisfy the
constraints from parity invariance \cite{Roiban:2004yf,
Witten:2004cp}. It is quite interesting to generalize this proof
in the theories with deformations, first at the linear order of
the deformation parameters.
\section*{Acknowledgements}
We are thankful to the CMS of Zhejiang University where the current work was initiated, Edna Cheung for organizing a great workshop, and to Konstantinos Zoubos for very helpful discussions. PG thanks AEI Potsdam, ICTP Trieste and the 4th Simons Workshop in Stony Brook for hospitality during
the course of this work. JW would also like to thank Bin Chen, Bo Feng, Bo-Yu Hou, Miao Li, Xing-Chang Song, Xiang Tang, Peng Zhang and Chuan-Jie Zhu for useful discussions, and Peking University,
Northwest University for hospitality. The work of PG was supported in part by the DOE grant DE-FG02-95ER40893. The work of JW is supported in part by the European Community's Human Potential
Programme under contract MRTN- CT-2004-005104 `Constituents, fundamental forces and symmetries of the universe' as a postdoc of the node of Padova.

\end{document}